\documentclass[onecolumn,prl,superscriptaddress]{revtex4}
\usepackage{color}
\usepackage[pdftex]{graphicx}
\usepackage{threeparttable}
\usepackage{amsmath,amsfonts,amssymb}
\usepackage[T1]{fontenc}
\newcommand{\be}{\begin{equation}}
\newcommand{\ee}{\end{equation}}

\begin{document}
\title{Proposal for Gravitational-Wave Detection Beyond the Standard Quantum Limit via EPR Entanglement}
\author{Yiqiu Ma}
\affiliation{Theoretical Astrophysics 350-17, California Institute of Technology, Pasadena, CA 91125, USA}
\author{Haixing Miao}
\affiliation{School of Physics and Astronomy, University of Birmingham, Birmingham, B15 2TT, United Kingdom}
\author{Belinda Heyun Pang}
\affiliation{Theoretical Astrophysics 350-17, California Institute of Technology, Pasadena, CA 91125, USA}
\author{Matthew Evans}
\affiliation{Massachusetts Institute of Technology, Cambridge, Massachusetts 02139, USA}
\author{Chunnong Zhao}
\affiliation{School of Physics, University of Western Australia, Western Australia 6009, Australia}
\author{Jan Harms}
\affiliation{Universit\`a degli Studi di Urbino ``Carlo Bo'', I-61029 Urbino, Italy}
\affiliation{INFN, Sezione di Firenze, Firenze 50019, Italy}
\author{Roman Schnabel}
\affiliation{Institut f{\"ur} Laserphysik and Zentrum f{\"ur} Optische Quantentechnologien, Universit{\"a}t Hamburg, Luruper Chaussee 149, 22761 Hamburg, Germany}
\author{Yanbei Chen}
\affiliation{Theoretical Astrophysics 350-17, California Institute of Technology, Pasadena, CA 91125, USA}
\date{\today}

\begin{abstract}
The Standard Quantum Limit in continuous monitoring of a system is given by the trade-off of shot noise and back-action noise. In gravitational-wave detectors, such as Advanced LIGO, both contributions can simultaneously be squeezed in a broad frequency band by injecting a spectrum of squeezed vacuum states with a frequency-dependent squeeze angle. This approach requires setting up an additional long base-line, low-loss filter cavity in a vacuum system at the detector's site. Here, we show that the need for such a filter cavity can be eliminated, by exploiting EPR-entangled signal and idler beams. By harnessing their mutual quantum correlations and the difference in the way each beam propagates in the interferometer, we can engineer the input signal beam to have the appropriate frequency dependent conditional squeezing once the out-going idler beam is detected. Our proposal is appropriate for all future gravitational-wave detectors for achieving sensitivities beyond the Standard Quantum Limit.
\end{abstract}

\maketitle

\noindent{\it Background and Summary.--}
Detection of gravitational waves from merging binary black holes (BBH) by the Laser Interferometer Gravitational-wave Observatory (LIGO) opened the era of gravitational wave astronomy\,\cite{GW150914}.  The future growth of the field relies on the improvement of detector sensitivity, and the vision for ground-based gravitational-wave detection is to improve, eventually by a factor $\sim$30 in amplitude in the next 30 years\,\cite{Punturo2010,Danilishin2012,Miao2014,Dwyer2015,GWsensitivity}. This will eventually allow us to observe all BBH mergers that take place in the universe, thereby inform on the formation mechanism of BBH, the evolution of the universe\,\cite{Dwyer2015,Sathyaprakash2012}, and the way gravitational waves propagate through the universe\,\cite{Tso2016,Kostelecky2016}. Higher signal-to-noise ratio observations of BBH will allow demonstrations and tests of effects of general relativity in the strong gravity and nonlinear regimes\,\cite{Lasky2016,Berti2013}. Besides BBH, gravitational waves from neutron stars are being highly anticipated, as well as an active program of joint EM-GW observations\,\cite{Chu2016,Metzger2012}.  Finally, improved sensitivity may lead to detections of more exotic sources\,\cite{cosmicstring2014}, as well as surprises.

A key toward better detector sensitivity is to suppress {\it quantum noise}, which arises from the quantum nature of light and the mirrors, and is driven by vacuum fluctuations of the optical  field entering from the dark port of the interferometer~\cite{Drever1976,Braginsky1977,Weiss1979,Caves1981}.  There are two types of quantum noise: shot noise, the finite displacement resolution due to the finite number of photons, and the radiation-pressure noise, which arises from the photons randomly impinging on the mirrors.  In the broadband configuration of Advanced LIGO, we measure the phase quadrature of the carrier field at the dark port, the quadrature that contains GW signal.  In this case, shot noise is driven by phase fluctuations of the incoming optical field, while radiation-pressure noise is driven by amplitude fluctuations. The trade off between these two types of noise, as dictated by the Heisenberg Uncertainty Principle,  gives rise to a sensitivity limitation called the Standard Quantum Limit (SQL)~\cite{Kimble2001,Buonanno2001,Buonanno2003}.

\begin{figure}[t] 
   \centering
   \includegraphics[width=4.25in]{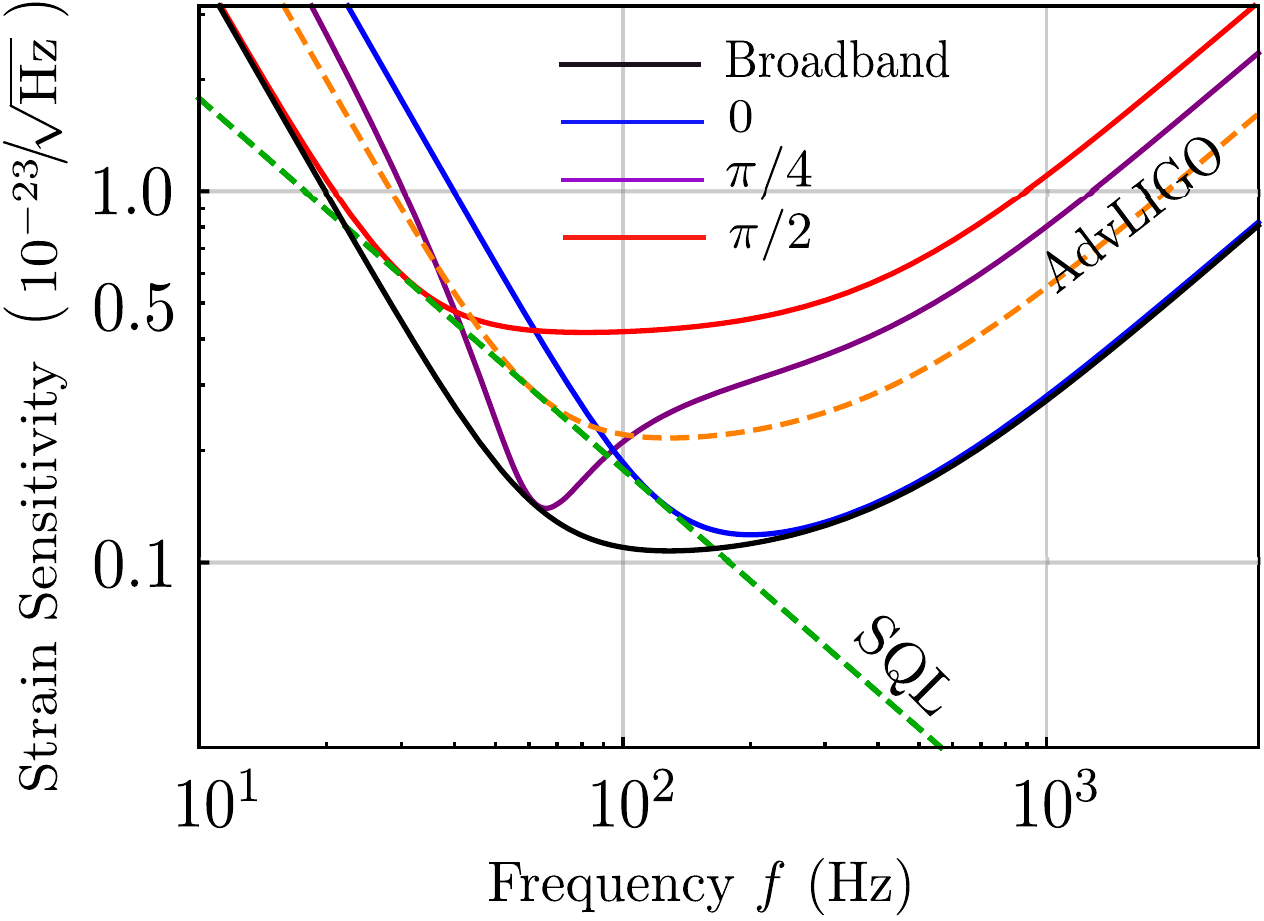} 
   \caption{Sensitivity of a AdvLIGO type gravitational wave detector driven by squeezed vacuum (6\,dB squeeze degree is chosen for comparing the $5\%$ input/output loss case in Fig.\,\ref{fig:sensitivity_curve}) with a fixed squeezing angle. The red, purple and blue curves describe the cases when squeezed light is injected with squeezing angle $0$(corresponding to an amplitude squeezed vacuum), $\pi/4$ and $\pi/2$, respectively. The black curve is the case when there is a frequency dependent squeezed vacuum injection. \label{fig:narrowband}}
\end{figure}

One way to improve LIGO's sensitivity with minimal modification to its optical configuration is to inject squeezed vacuum into the dark port\,\cite{Caves1981,Unruh1982,Schnabel2010,LIGOwhitepaper}. 
More than 10\,dB of squeezing down to audio side-band frequency (10\,Hz to 10\,kHz) has been demonstrated  in the lab~\cite{Vahlbruch2008,Mehmet2011,Chua2011,Stefszky2012,McKenzie2004,Vahlbruch2016}, while moderate noise reductions have  been demonstrated in the large-scale interferometers GEO\,600~\cite{squeezingLIGO2011} and LIGO~\cite{squeezingLIGO2013}. However, squeezed vacuum generated by a nonlinear crystal via Optical Parametric Amplification (OPA) is frequency-independent for audio sidebands: within the GW band, we can only ``squeeze''  a fixed quadrature --- fluctuations in the orthogonal quadrature are amplified by the same factor, as required by the Heisenberg Uncertainty Principle.  This does not allow broadband improvement of  sensitivity beyond the SQL~\cite{Jaekel1990,Kimble2001} such as the example shown in Fig.\,\ref{fig:narrowband}; instead,   
a frequency-dependent quadrature must be squeezed for each sideband frequency.  Starting off from frequency-independent squeezing, we must {\it rotate} the squeezed quadrature in a frequency-dependent way~\cite{Kimble2001,Chelkowski2005}; for the broadband configuration of Advanced LIGO, this rotation angle needs to gradually transition by $\pi/2$ at a frequency scale of 50\,Hz~\cite{Evans2013}.  Kimble {\it et al.}\,\cite{Kimble2001} proposed to achieve such rotation by filtering the field with two Fabry-Perot cavities; Khalili further showed that it is often sufficient to use one cavity with bandwidth and detuning (from the carrier frequency) roughly at the transition frequency\,\cite{Khalili2007,Khalili2010}.   However, the narrowness of the bandwidth requires the filter cavity to be long in order to limit impact from optical losses; the current plan for Advanced LIGO is to construct a $\sim 16$\,m filter cavity~\cite{Isogai2013,Evans2013,Kwee2014}, and $\sim 300$m long cavities have been studied for KAGRA\,\cite{Caposcasa2016} and for the Einstein Telecscope\,\cite{ET2011}.  Alternative theoretical proposals for creating narrowband filter cavities were also discussed, they are strongly limited by thermal noise and/or optical losses\,\cite{Mikhailov2006,Ma2014,Qin2014}.

In this paper, we propose a novel strategy to achieve broadband squeezing of the total quantum noise via the preparation of EPR entanglement and the dual use of the interferometer as both the GW detector and the filter, eliminating the need for external filter cavities.

\begin{figure}[t] 
   \centering
   \includegraphics[width=4.5in]{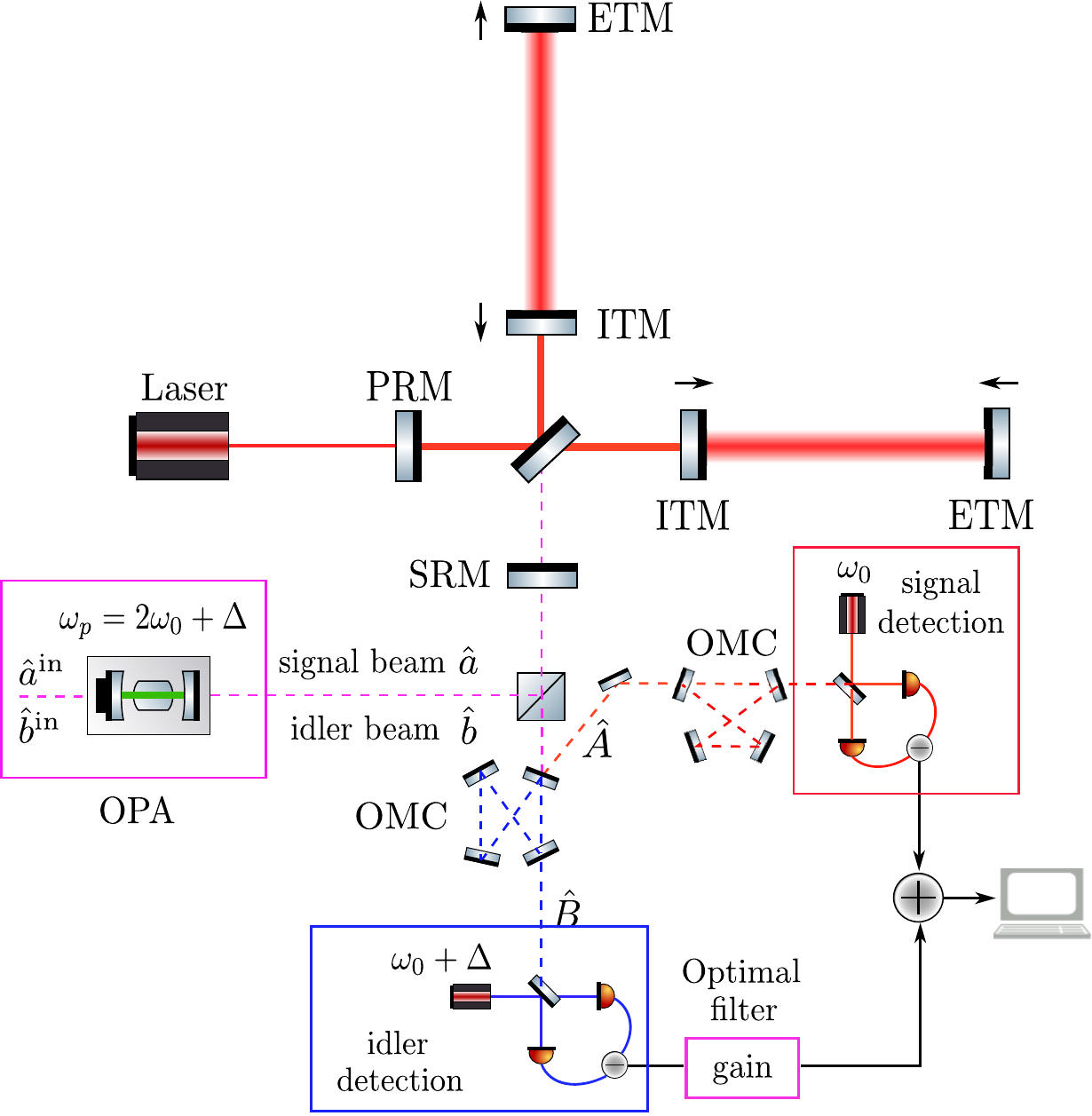} 
   \caption{Optical configuration for  noise suppression via EPR entanglement. The OPA is detuned by $\Delta$, generating signal beam $a$ and idler beam $b$ and injecting them into the interferometer.  Upon returning from the interferometer, signal beam $A$ and idler beam $B$ are separated and filtered by the output mode cleaners (denoted as OMC in the figure), and each detected via homodyne detection. Measurement data are combined using an optimal filter for obtaining the squeezing of the quantum noise on the signal channel. The abbreviations PRM, ITM, ETM and SRM stand for power recycling mirror, input test mass mirror, end test mass mirror and signal recycling mirror, respectively.
   }
   \label{fig:scheme2}
\end{figure}

\begin{figure}[t] 
   \centering
   \includegraphics[width=4.25in]{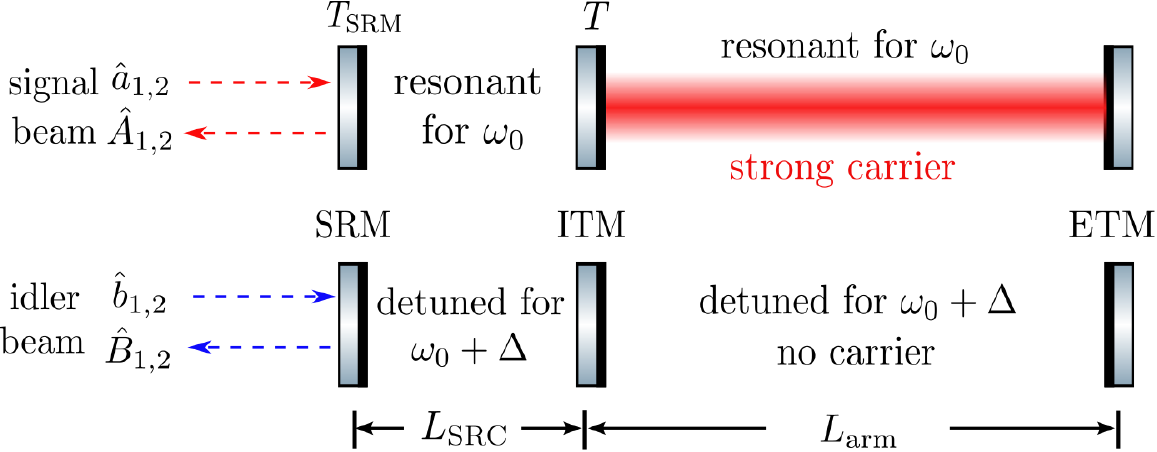} 
   \caption{The differential mode of the interferometer as seen by the signal (upper panel) and idler (lower panel) beams.   \label{fig:IFO}}
\end{figure}

As shown in Fig.~\ref{fig:scheme2}, our strategy is divided into 4 steps. (i) We detune the pumping frequency of the OPA away from $2\omega_0$ (where $\omega_0$ is the carrier frequency of the interferometer) to $\omega_p = 2\omega_0+\Delta$,  with $\Delta$ an rf frequency of a few MHz,  creating two EPR-entangled beams: the {\it signal beam} around the carrier frequency $\omega_0$,  and the {\it idler beam} around $\omega_0 +\Delta$. (ii) The idler beam, being far detuned from the carrier, sees the interferometer as a simple detuned cavity, and experiences frequency-dependent quadrature rotation, see Fig.~\ref{fig:IFO}, which can be optimised by adjusting $\Delta$ with respect to the lengths of interferometer cavities. (iii) 

\begin{figure}[t] 
   \centering
   \includegraphics[width=4.4in]{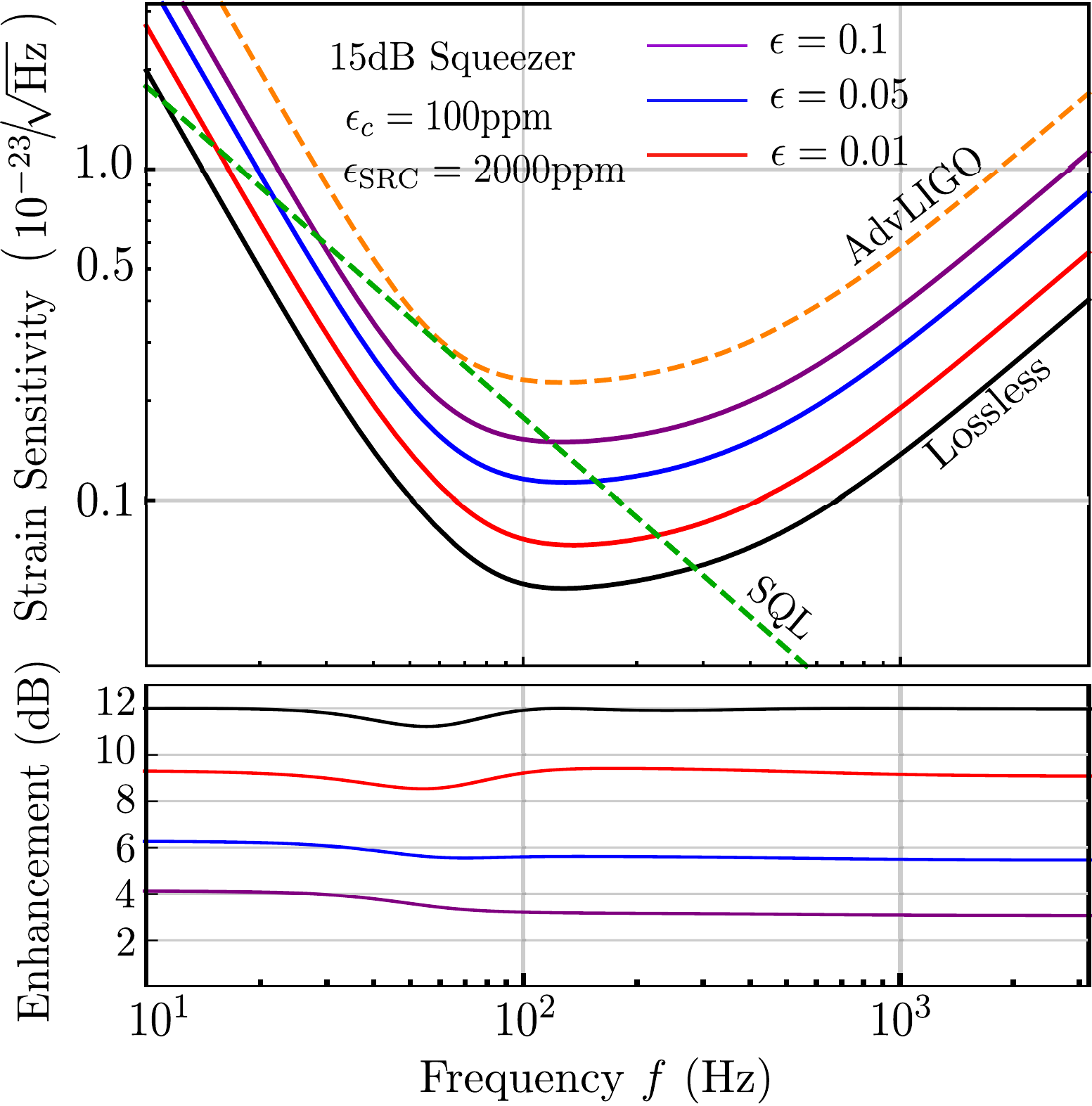} 
   \caption{Upper panel: Noise spectrum of Advanced LIGO configurations with conditional frequency-dependent squeezing by using a 15\,dB squeezed vacuum at MHz frequencies(see Table~\ref{tab:parameters}), assuming no loss (black), and assuming arm-cavity loss $\epsilon_c=100\,$ppm and signal recycling cavity loss $\epsilon_{\rm SRC}=2000\,$ppm, plus an identical input and output loss  $\epsilon$ of 1\% (red), 5\% (blue) and 10\% (purple). Lower panel: The sensitivity improvement factor measured in terms of dB.
      \label{fig:sensitivity_curve}}
\end{figure}

When traveling out of the interferometer, the collinear signal and idler beams are separated and filtered by the output mode cleaners and measured by beating with local oscillators at frequencies $\omega_0$ and $\omega_0+\Delta$, respectively.  (iv) The homodyne measurement of  a fixed quadrature of the out-going idler beam  conditionally squeezes the {\it input signal beam} in a frequency dependent way, thereby achieving the broadband reduction of quantum noise. Practically, benefit of the conditional squeezing of the signal beam is obtained as we apply a Wiener filter to the photocurrent of the idler and subtract it from the photocurrent of the signal beam. Without optical losses, using parameters  in Table\,\ref{tab:parameters} (with a  15\,dB squeezed vacuum in particular), we obtain the solid black curve in Fig.~\ref{fig:sensitivity_curve}, with $\sim$11-12\,dB improvement over the entire frequency band. We shall  next discuss more details of the configuration, as well as the impact of optical losses; further details are provided as Supplementary Materials. 

\begin{table}[t]
   \centering
   \begin{tabular}{ |l|l|l| }
   \hline 
  $\lambda$ & Carrier laser wavelength &1064\,nm\\
  $T_{\rm SRM}$ & SRM power transmissivity & 0.35\\
   $T$ & ITM power transmissivity & 0.014\\
  $L_{\rm arm}$ & Arm cavity length & $\sim 4$\,km\\
  $L_{\rm SRC}$& Signal recycling cavity length &$\sim 50$\,m\\
  $\gamma$ & Detection bandwidth & 389\,Hz\\
  $m$ & Mirror mass (ITM and ETM) & 40\,kg\\
  $I_c$ & Intra cavity power & 650\,kW\\
  $\Delta$ & 
  Idler-signal detuning
  & $-15.3$\,MHz\\
  $r$ &Squeeze factor of the OPA& 1.23 (15\, dB)\\
  \hline
  \end{tabular}
  \caption{Sample Parameters for Advanced LIGO. (See supplementary material for details.)}
   \label{tab:parameters}
\end{table}

\begin{figure}[t] 
   \centering
   \includegraphics[width=4.5in]{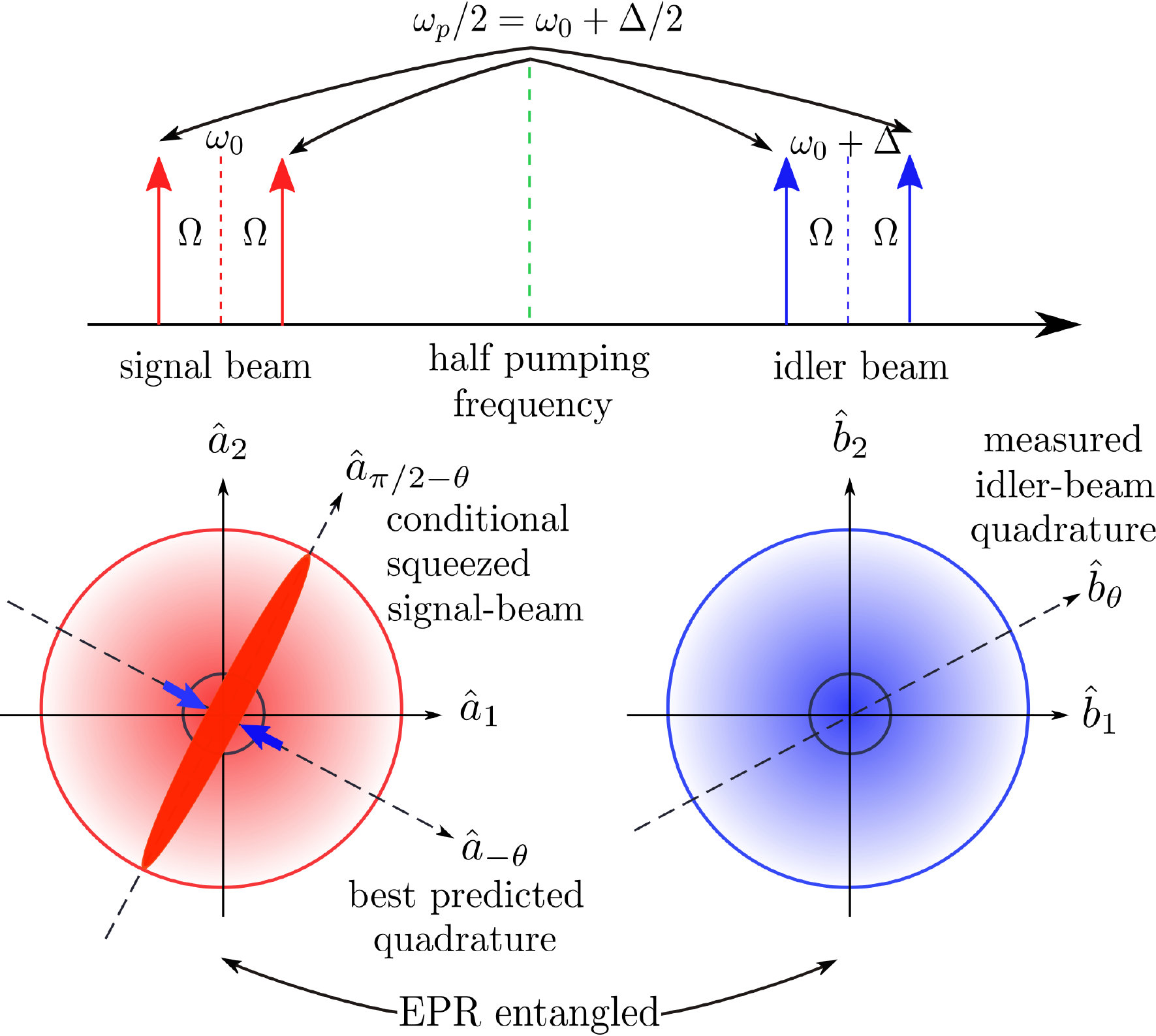} 
   \caption{Spectral decomposition of EPR-entangled beams (upper panel) and the quantum statics of the signal and idler beams (lower panel).
   \label{fig:EPR}}
\end{figure}

\noindent{\it EPR entanglement by detuning the OPA.--}   

For an OPA pumped at $\omega_p$, it is often convenient to study {\it quadrature fields} around $\omega_p/2$, which are linear combinations of upper and lower sideband fields at $\omega_p/2 \pm\Omega$, with $\zeta$ quadrature defined by:
\begin{equation}
\hat c_\zeta(\Omega) =\left(e^{-i\zeta} \hat c_{\omega_p/2 + \Omega} +e^{+i\zeta} \hat c_{\omega_p/2-\Omega}^\dag\right)/\sqrt{2}
\end{equation}
Here $\hat c_\omega$ and $\hat c^\dagger_\omega$ are the annihilation and creation operators for the optical field at $\omega$; we will use $\hat c_{1,2}$ to stand for $\hat c_{0,\pi/2}$, and $\hat c_\zeta= \hat c_1\cos\zeta+\hat c_2\sin\zeta$.  For a  squeeze factor $r$ and squeeze angle $\theta$, the orthogonal quadratures $\hat c_\theta$ and $\hat c_{\theta+\pi/2}$ have uncorrelated fluctuations, with spectra given by 
\begin{equation}
S_{\hat c_\theta \hat c_\theta}=e^{-2r}\,, \;
S_{\hat c_{\theta+{\pi/2}} \hat c_{\theta+{\pi/2}}}=e^{+2r}\,.
\end{equation} 
Compared with vacuum, fluctuations in $\hat c_\theta$ are suppressed by $e^{2r}$, and those in $\hat c_{\theta+\pi/2}$ are amplified by $e^{2r}$. This is due to the entanglement between the upper and lower sidebands, $\omega_p/2\pm\Omega$, generated by the optical nonlinearity. However,  any pair of sideband fields with frequencies $\omega_1$ and $\omega_2$ within the squeeze bandwidth (usually $>$MHz) from $\omega_p/2$, and satisfying $\omega_1+\omega_2=\omega_p$, are entangled; in particular, for the proposed OPA (Fig.\,\ref{fig:scheme2}) with pumping frequency $\omega_p = 2\omega_0+\Delta$, we have entanglement between $\omega_0 + \Omega$ and $\omega_0 +\Delta - \Omega$, as well as $\omega_0 - \Omega$ and $\omega_0 +\Delta + \Omega$, as shown in the upper panel of Fig.~\ref{fig:EPR}. As it turns out, this entanglement is equivalent to an EPR-type entanglement\,\cite{Zhang2003,Marino2007,Hage2010}  between quadratures around $\omega_0$ [consisting of $\omega_0\pm\Omega$ sidebands, denoted by $\hat a_{\zeta}(\Omega)$] and those around $\omega_0+\Delta$ [consisting $\omega_0+\Delta\pm\Omega$ sidebands, denoted by $\hat b_{\zeta}(\Omega)$].  In terms of the four fields, $(\hat a_{1}\pm \hat b_1 )/\sqrt{2}$ and $(\hat a_{2}\pm\hat  b_2 )/\sqrt{2}$, they are mutually uncorrelated, and have spectra
\begin{equation}
S_{(\hat a_1\pm \hat b_1)/{\sqrt{2}}}= e^{\pm 2r}\,,\quad  S_{({\hat a_2\pm\hat  b_2})/{\sqrt{2}}}= e^{\mp 2r}\,.
\end{equation}
In other words,  for $r\stackrel{>}{_\sim}1$, fluctuations in $\hat b_1 - \hat  a_1$ and  $\hat b_2 + \hat a_2$ are both much below vacuum level, as in the  original EPR situation. In this way (lower panel of Fig.~\ref{fig:EPR}), if we detect $\hat b_\theta =\hat b_1\cos\theta+\hat b_2\sin\theta$, we can predict $\hat a_{-\theta}= \hat a_1\cos\theta - \hat a_2\sin\theta$ with a very good accuracy, while not providing any information for $\hat a_{\pi/2-\theta}$.  More precisely, given measurement data of the idler quadrature $\hat b_\theta$,  the signal beam will be conditionally squeezed, with  {\it conditional spectra}
\begin{equation}
S^{|\hat b_\theta}_{\hat a_{-\theta}\hat a_{-\theta}}  = {1}/{\cosh (2r)}\,,\;
S^{|\hat b_\theta}_{\hat a_{\frac{\pi}{2}-\theta}\hat a_{\frac{\pi}{2}-\theta}}  = \cosh (2r)\,,
\end{equation}
where the squeeze angle is $-\theta$, and the squeeze factor is $(\log\,\cosh(2r))/2$.  For significant squeezing, $e^{2r} \gg 1$, this corresponds to 3\,dB less squeezing than before detuning the pump field.

\noindent{\it Improvement of Detector Sensitivity.--}  As shown in Fig.1,  after signal beam $\hat a_{1,2}$ and idler beam $\hat  b_{1,2}$ are fed into the interferometer, we detect phase quadratures of the out-going signal and the idler beams, $A_2$ and $B_2$, after they are separated and filtered by the output mode cleaners (Fig.~\ref{fig:scheme2}).  For the signal beam (upper panel of Fig.~\ref{fig:IFO}), we have\,\cite{Kimble2001}:
\begin{equation}
\hat A_2 = e^{2i\beta} (\hat a_2 - \mathcal{K} \hat a_1) +\sqrt{2\mathcal{K}}e^{i\beta} h/h_{\rm SQL}\,,
\end{equation}
which consists of shot noise, radiation-pressure noise, and signal, with  $\beta=\arctan(\Omega/\gamma)$, where $\gamma$ is the bandwidth of the interferometer seen by the signal beam, 
\begin{equation}
h^2_{\rm SQL}={8\hbar}/{(m\Omega^2L^2)},\quad
\mathcal{K}=2\Theta^3\gamma/[\Omega^2(\Omega^2+\gamma^2)],
\end{equation}
and $\Theta =[8\omega_0 I_c/(m L c)]^{1/3}$. 

Here we need to squeeze the $a_{[-\arctan(1/\mathcal{K})]}$ quadrature of the input signal beam, which requires detecting  $b_{\arctan(1/\mathcal{K})}$.  If we detect $B_2$, we will need the interferometer (lower panel of Fig.~\ref{fig:IFO}) to apply a rotation of 
$
\Phi_{\rm rot} = \arctan \mathcal{K}
$
to the idler beam so that $\hat B_2=\hat b_{\arctan(1/\mathcal{K})}$.  This can be realized approximately by adjusting the detuning $\Delta$ and the length of signal-recycling cavity and arm cavity (see Supplementary Material for details),  if  $\Theta \ll \gamma$.  To achieve the sensitivity provided by conditional squeezing, we need to compute the best estimate of $\hat A_2$ from $\hat B_2$, and subtract it from $\hat A_2$.  If a rotation by $\Phi_{\rm rot}$ is realized exactly, we will have a noise spectrum of
\be
S_{h}\approx \frac{h_{\rm SQL}^2}{2\cosh{2r}}\left(\mathcal{K}+\frac{1}{\mathcal{K}}\right),
\ee
where conditional squeezing provides a $\cosh\,2r$ suppression. In reality, we get less suppression since the interferometer, acting as a single cavity, does not exactly realize $\Phi_{\rm rot}$ for the idler beam. In Fig.~\ref{fig:sensitivity_curve}, the black curve shows the actual noise spectrum for parameters in Table~\ref{tab:parameters}.

\noindent {\it Discussions.--} In Fig.~\ref{fig:sensitivity_curve}, we plot noise spectra of interferometers with optical losses.  In particular, we include losses in the arm cavities, at the input port, and during readout.  As it turns out, the current 100\,ppm arm cavity loss and 2000\,ppm has only a small effect on the noise (for details, see Supplementary Material).  When the input loss and the readout loss are both around $10\%$, the sensitivity improvement is only roughly $3$\,dB, which corresponds to an amplitude improvement $\sim 1.4$. However, for a lower loss of 5\%, which is promising in the near future\,\cite{Evans2013,Barsotti2016,Barsotti2014}, we can gain 6\,dB or a factor of $\sim 2$ improvement in amplitude. This corresponds to an increase of sensitive sky volume by a factor of 8. Compared to the traditional scheme with a filter cavity\,\cite{Evans2013}, our input and detection losses are doubled, because signal and idler beams experience the same amount of loss during propagation.  Although we do suffer less from loss in the filter cavity compare to the design based on an auxiliary filter cavity (since arm cavities have less loss), this higher level of input and detection losses is the price we have to pay in this scheme for eliminating the additional filter cavity.

\noindent{\it Acknowledgements.--} Research of YM, BHP, YC is supported by NSF grant PHY-1404569 and PHY-1506453, as well as the Institute for Quantum Information and Matter, a Physics Frontier Center. H.M. is supported by the Marie-Curie Fellowship and UK STFC Ernest Rutherford Fellowship. CZ would like to thank the support of Australian Research Council Discovery Project DP120104676 and DP120100898. RS is supported by DFG grant SCHN757/6 and by ERC grant 339897 (`Mass Q')

\bibliographystyle{unsrt}

\section{Supplementary Material}
\section{Derivation of the sensitivity formula}
First, for each audio-sideband frequency $\Omega$, the field input-output relations of the squeezer (the pumped OPA) can be written as:
\begin{equation}\label{eq:crystal_inout_sideband}
\begin{split}
&\hat a(\omega_0+\Omega)=\mu \hat a_{\rm in}(\omega_0+\Omega)+\nu\hat b^\dag_{\rm in}(\omega_0+\Delta-\Omega),\quad \hat b(\omega_0+\Delta+\Omega)=\mu \hat b_{\rm in}(\omega_0+\Delta+\Omega)+\nu\hat a^\dag_{\rm in}(\omega_0-\Omega);\\
&\hat a^\dag(\omega_0-\Omega)=\mu^* \hat a^\dag_{\rm in}(\omega_0-\Omega)+\nu^*\hat b_{\rm in}(\omega_0+\Delta+\Omega),\quad\hat b^\dag(\omega_0+\Delta-\Omega)=\mu^* \hat b^\dag_{\rm in}(\omega_0+\Delta-\Omega)+\nu^*\hat a_{\rm in}(\omega_0+\Omega),
\end{split}
\end{equation}
where $\hat a$ and $\hat b$ describe the generated signal and idler fields near $\omega_0$ and $\omega_0\pm\Delta$, respectively.  The fields $\hat a_{\rm in},\hat b_{\rm in}$ represent the vacuum fields entering into the squeezer. The phenomenological coefficient $\mu$ and $\nu$ are determined by the $\chi^{(2)}-$nonlinearity coefficient of the crystal and the pumping field strength\,\cite{Scully1997}. Field commutation relation requires them to satisfy the relation $|\mu^2|-|\nu|^2=1$. Since the phase of $\mu$ and $\nu$ can be absorbed into the definition of creation and annihilation operators, we can parametrise them as $\mu=\cosh{r}$ and $\nu=\sinh{r}$, where $r$ is usually denoted to be the squeezing degree of the OPA. In the so-called  two-photon formalism where we define:
\be\label{eq:quadratures}
\begin{split}
&\hat a_{1}(\Omega)=\frac{\hat a(\omega_0+\Omega)+\hat a^\dag(\omega_0-\Omega)}{\sqrt{2}},\quad \hat b_{1}(\Omega)=\frac{\hat b(\omega_0+\Delta+\Omega)+\hat b^\dag(\omega_0+\Delta-\Omega)}{\sqrt{2}};\\
&\hat a_{2}(\Omega)=\frac{\hat a(\omega_0+\Omega)-\hat a^\dag(\omega_0-\Omega)}{\sqrt{2}i},\quad \hat b_{2}(\Omega)=\frac{\hat b(\omega_0+\Delta+\Omega)-\hat b^\dag(\omega_0+\Delta-\Omega)}{\sqrt{2}i},\\
\end{split}
\ee
the relations in Eq.\,\eqref{eq:crystal_inout_sideband} then can be represented in another form (in the following,  $\hat a_{1,2}(\Omega)$ and $\hat b_{1,2}(\Omega)$ will be simply written as $\hat a_{1,2}$ and $\hat b_{1,2}$):
\begin{align}\label{eq:crystal_inout}
&\hat a_1+\hat b_1=e^r(\hat a_{\rm in 1}+\hat b_{\rm in1}),\quad \hat a_1-\hat b_1=e^{-r}(\hat a_{\rm in 1}-\hat b_{\rm in1});\\
&\hat a_2+\hat b_2=e^{-r}(\hat a_{\rm in 2}+\hat b_{\rm in2}),\quad \hat a_2-\hat b_2=e^{r}(\hat a_{\rm in 2}-\hat b_{\rm in2}),
\end{align}
(the $\hat a_{\rm in 1,2}, \hat b_{\rm in 1,2}$ are defined in the same way as Eq.\,\eqref{eq:quadratures}).
EPR-type commutation relation $[\hat a_1-\hat b_1,\hat a_2+\hat b_2]=0$ allows the existence of the state in which the fluctuations of quadrature combinations $(\hat a_1-\hat b_1)/\sqrt{2}$ and $(\hat a_2+\hat b_2)/\sqrt{2}$ are much below the vacuum level. Therefore $\hat b_1$ is correlated with $\hat a_1$ while $\hat b_2$ is correlated with $-\hat a_2$. Therefore $\hat a_{-\theta}=\hat a_1\cos\theta-\hat a_2\sin\theta$ correlates with $\hat b_{\theta}=\hat b_1\cos\theta+\hat b_2\sin\theta$. When we do conditioning by combining the measurement results of signal and idler fields, we assume the measurement result of the idler field quadrature $\hat b_\theta$ is filtered with a filtering gain factor $g$ and then combined with the signal field quadrature $\hat a_{-\theta}$, leads to:
\be
\hat a^g_{-\theta}=\hat a_{-\theta}-g\hat b_\theta=(\hat a_1-g\hat b_1)\cos\theta-(\hat a_2+g\hat b_2)\sin\theta.
\ee
It is easy to show that the spectrum of $\hat a^g_{-\theta}$ is:
\be
S_{\hat a^g_{-\theta}\hat a^g_{-\theta}}=(\mu-g\nu)^2+(\nu-g\mu)^2.
\ee
For realising an optimal filtering (which is the so-called ``Wiener filtering") so that $S_{\hat a^g_{-\theta}\hat a^g_{-\theta}}$ takes its minimum value, we can solve $\delta S_{\hat a^g_{-\theta}\hat a^g_{-\theta}}/\delta g=0$, which leads to the Wiener filter gain factor $g_{\rm opt}$ and conditional squeezing spectrum:
\be
g_{\rm opt}=\frac{2\mu\nu}{\mu^2+\nu^2}=\tanh{2r},\quad S^{\rm cond}_{\hat a_{-\theta}\hat a_{-\theta}}=\frac{1}{\mu^2+\nu^2}=\frac{1}{\cosh{2r}}.
\ee

In laser interferometer gravitational wave detectors, we have the input-output relation for quantum noise field in the signal channel as\,\cite{Kimble2001}:
\begin{equation}\label{eq:signal_in-out}
\begin{split}
&\hat A_2 = e^{2i\beta} (\hat a_2 - \mathcal{K} \hat a_1)=e^{2i\beta} (\sqrt{1+\mathcal{K}^2})(\hat a_1\cos\xi-\hat a_2\sin\xi),\\
\end{split}
\end{equation}
where $\hat A_2$ is the phase quadrature of the signal fields propagate out of the interferometer and $\xi=-\arctan{1/\mathcal{K}}$. If we want the phase quadrature of the idler fields out of the interferometer $\hat B_2$ to maximally correlate with $\hat A_2$ in Eq.\,\eqref{eq:signal_in-out}, then $\hat B_2=\hat b_{\rm \arctan{1/\mathcal{K}}}$ (besides an unimportant phase factor $\alpha$ accumulated by sidebands of the idler field during its propagation ). Therefore, the rotation angle of the idler field $\Phi_{\rm rot}$ by the interferometer defined in
\be
\hat B_2=e^{i\alpha}(-\hat b_{1}\sin\Phi_{\rm rot}+\hat b_2\cos\Phi_{\rm rot}),
\ee
is given as $\Phi_{\rm rot}=\arctan{\mathcal{K}}$.

Similarly, when combing the measurement results of signal and idler channel, we have:
\be
\hat A^g_2=\hat A_2-g\hat B_2\quad{\rm and}\quad S_{\hat A^g_2\hat A^g_2}=S_{\hat A_2\hat A_2}+|g|^2
S_{\hat B_2\hat B_2}-g^*S_{\hat A_2\hat B_2}-gS_{\hat B_2\hat A_2}.
\ee
Variation with respect to filter gain factor $g$ leads to the Wiener filter and minimum variance given as:
\begin{align}\label{eq:opt_var}
&g_{\rm opt}=\frac{S_{\hat A_2\hat B_2}}{S_{\hat B_2\hat B_2}}=e^{i(2\beta-\alpha)}\sqrt{1+\mathcal{K}^2}\tanh{2r},\\
&S^{\rm cond}_{\hat A_2\hat A_2}=S_{\hat A_2\hat A_2}-\frac{S_{\hat B_2\hat A_2}S_{\hat A_2\hat B_2}}{S_{\hat B_2\hat B_2}}=\frac{1+\mathcal{K}^2}{\cosh{2r}}.
\end{align}
Considering the signal field as: $\hat A_2^{\rm GW}=\sqrt{2\mathcal{K}}e^{i\beta}h/h_{\rm SQL}$\,\cite{Kimble2001}, we can recover the Eq.\,(7) of the main text:
\be
S_{\rm hh}
=\frac{h_{\rm SQL}^2}{2\cosh{2r}}\left(\mathcal{K}+\frac{1}{\mathcal{K}}\right).
\ee

\section{Parameter setting}
\subsection{Requirements}
Our conditional squeezing scheme is based on the cancelation of the results from signal beam detection and idler beam detection. If the squeezer's squeezing level is high, the parameter error will have a significant effect on the final squeezing level. For example,  the effect of variation of the idler rotation angle to the sensitivity is roughly given by:
\be\label{eq:parameter_tolerance}
S_{\rm hh}=\frac{h_{\rm SQL}^2}{2\cosh{2r}}\left(\mathcal{K}+\frac{1}{\mathcal{K}}\right)
+\frac{h_{\rm SQL}^2}{2}\frac{(\sinh{2r})^2}{\cosh{2r}}\left(\mathcal{K}+\frac{1}{\mathcal{K}}\right)\delta\Phi^2.
\ee
For a 15\,dB squeezer as shown in the main text, the ratio between the correction term and the exact value is roughly $\approx 249\delta\Phi^2$. Therefore even a $10\%$ relative correction to the noise spectrum requires the error of the rotation angle to be as small as $0.02$\,rad. This simple estimation tells us that it is of great importance to search the suitable parameters for our proposed scheme.

In our design, the signal field sees an interferometer working in the resonant sideband extraction mode while the idler field sees the interferometer as a filter cavity. This filter cavity should rotate the idler field in its phase space by an angle $\Phi_{\rm rot}=\arctan{\mathcal{K}}$. Generally, for realising such a rotation angle, two filter cavities are required\,\cite{Kimble2001} (for a more general discussion, see\,\cite{Purdue2002}). However, when the signal field works in the resonant sideband extraction mode, the interferometer has a relatively large bandwidth so that $\mathcal{K}$ can be approximated around the transition frequency as: $\mathcal{K}\approx2\Theta^3/(\Omega^2\gamma)$. In this case, only one filter cavity is necessary to achieve the required rotation of the idler field, and we use the signal recycling interferometer itself as the filter. The required bandwidth and detuning of the signal recycling interferometer withe respect to the idler field is given by\,\cite{Purdue2002,Khalili2007}:
\be\label{eq:fit_condition}
\gamma_f = \sqrt{\Theta^3/\gamma},\,\quad\delta_f=-\gamma_f.
\ee
\subsection{Parameter setting conditions}
The dependence of detuning $\delta_f$ and bandwidth $\gamma_f$ on the interferometer parameters can be seen in the interferometer reflectivity, which is given by (in the sideband picture)\,\cite{Buonanno2003}:
\be
\begin{split}\label{eq:idler_reflectivity}
r^{\rm idler}_{\rm ifo}(\Omega)=\frac{\rho+(\tau\tilde{\tau}-\rho\tilde{\rho}){\rm exp}[2i(\Delta+\Omega)L_{\rm arm}/c]}{1-\tilde{\rho} {\rm exp}[2i(\Delta+\Omega)L_{\rm arm}/c]},;\\
\end{split}
\ee
where $\rho$, $\tilde{\rho}$, $\tau$, and $\tilde{\tau}$ describe the reflectivity and transmissivity of the signal recycling cavity. They are given by\,\cite{Buonanno2003}:
\be\label{eq:srcreflectiontransmission}
\begin{split}
&\tilde{\rho}=\frac{\sqrt{R_{\rm ITM}}-\sqrt{R_{\rm SRM}}{\rm exp}[2i\phi_{\rm SRC}]}{1-\sqrt{R_{\rm ITM}R_{\rm SRM}}{\rm exp}[2i\phi_{\rm SRC}]},\quad \rho=-\frac{\sqrt{R_{\rm SRM}}-\sqrt{R_{\rm ITM}}{\rm exp}[2i\phi_{\rm SRC}]}{1-\sqrt{R_{\rm ITM}R_{\rm SRM}}{\rm exp}[2i\phi_{\rm SRC}]},\\
&\qquad\qquad\qquad\tau=\tilde{\tau}=\frac{i\sqrt{T_{\rm SRM}}\sqrt{T_{\rm ITM}}{\rm exp}[i\phi_{\rm SRC}]}{(1-\sqrt{R_{\rm SRM}R_{\rm ITM}}{\rm exp}[2i\phi_{\rm SRC}])}.
\end{split}
\ee
Here, $R_{\rm ITM},R_{\rm SRM}$ are the power reflectivity of the input test mass mirrors and the signal recycling mirror and the $\phi_{\rm SRC}$ is the single trip phase of the idler field in the signal recycling cavity given as:
\be\label{eq:src_phase}
\phi_{\rm SRC}=\Delta L_{\rm SRC}/c.
\ee
From Eq.\eqref{eq:idler_reflectivity}, the resonance condition can be derived as:
\be\label{eq:resonance}
\boxed{
{\rm Mod}_{2\pi}\left[2(\delta_f+\Delta)\left(\frac{L_{\rm arm}}{c}\right)+{\rm Arg}[\tilde{\rho}]\right]=0,}
\ee
which determines the detuning $\delta_f$. 
The bandwidth $\gamma_f$ is given by:
\be\label{eq:idler_bandwidth}
\boxed{
\gamma_{f}\approx\frac{T_{\rm SRM}}{1+R_{\rm SRM}+2\sqrt{R_{\rm SRM}}\cos{2\phi_{\rm SRC}}}\gamma_{\rm ITM},}
\ee
where $\gamma_{\rm ITM}=cT_{\rm ITM}/(4L_{\rm arm})$.

From Eq.\eqref{eq:resonance} and \eqref{eq:idler_bandwidth}, when the reflectivity of the signal recycling mirror and the input test mass mirror is given, we have the following tunable parameters: (1) detuning of the idler with respect to the signal $\Delta$, (2) arm cavity length $L_{\rm arm}$, (3) the phase $\phi_{\rm SRC}$. These parameters must be tuned in such a way so that arm cavity and signal recycling cavity must each be resonant with the signal carrier frequency $\omega_0$ thereby the signal channel will not be affected. This means that the length tuning of the signal recycling cavity and the arm cavities (denoted by $\delta L_{\rm arm}$ and $\delta L_{\rm SRC}$, respectively) , starting from their initial lengths (denoted by $L^{(0)}_{\rm arm}$ and $L^{(0)}_{\rm SRC}$, respectively) should be integer numbers of half wavelength of the main carrier field, that is
\be
 \delta L_{\rm SRC}=p\lambda/2,\quad \delta L_{\rm arm}=q\lambda/2, \quad(p,q\in\mathbb{Z}).
\ee
Also note that Eq.\eqref{eq:fit_condition} tells us that $\gamma_f$ and $\delta_f$ depends on $\delta L_{\rm arm}$ while not on $\delta L_{\rm SRC}$. Since $L_{\rm arm}^{(0)}$ is typically of kilometer scale, thereby $\delta L_{\rm arm}$ as a small length tuning has negligible effect on the value of required $\gamma_f$ and $\delta_f$.

To obtain the required bandwidth $\gamma_f$, we can tune the $\phi_{\rm SRC}$ by tuning $\Delta$ and $L_{\rm SRC}$ in Eq.\eqref{eq:src_phase} to satisfy:
\be
\phi_{\rm SRC}=\Delta L_{\rm SRC}/c=\frac{1}{2}\arccos{\left[\frac{\gamma_{\rm ITM}}{\gamma_f}T_{\rm SRM}-(1+R_{\rm SRM})\right]}+n\pi,
\ee
or in another form:
\be
\Delta=\frac{c}{2L_{\rm SRC}}\arccos{\left[\frac{\gamma_{\rm ITM}}{\gamma_f}T_{\rm SRM}-(1+R_{\rm SRM})\right]}+\frac{n\pi c}{L_{\rm SRC}},
\ee
which tells us that $n$, as a tunable degree of freedom, represents how many free-spectral range of signal recycling cavity contained by the idler detuning $\Delta$.  In summary, we have three tunable integers: $m,n$ and $p$. For a fixed value of $n$, the phase $\phi_{\rm SRC}$ only depends on the $L_{\rm SRC}=L_{\rm SRC}^{(0)}+p\lambda/2$, thus the rough range of $p$ can be determined. To obtain the required detuning $\delta_f$, we need to further do fine tuning of $p$ and then do a corresponding tuning of $q$ to match the resonance condition Eq.\eqref{eq:resonance}. A sample parameter set is given in Tab.\,\ref{tab:parameters}. We also need to emphasise that the practical parameters setting should be decided considering concrete experimental requirements, and a feedback control system for length tuning needs to be carefully designed, what we have here is merely an example demonstrating that these parameters can in principle be found.

\subsection{Phase compensation}
Note that since the detuned $\hat b-$ field will pick up a phase when it is reflected by the signal recycling cavity, which will contribute an additional rotation angle, we need to compensate this phase by properly choosing the homodyne angle. This fact can be seen by manipulating\eqref{eq:idler_reflectivity}, given as follows.

From Eq. \eqref{eq:srcreflectiontransmission}, we can derive that:
\be
\frac{\tilde{\rho}^{*}}{\rho}=-\frac{1}{\tau^2-\rho\tilde{\rho}}.
\ee
Substituting this relation into Eq.\eqref{eq:idler_reflectivity} leads to:
\be
\begin{split}
r^{\rm idler}_{\rm ifo}(\Omega)&=\frac{-\tilde{\rho}^*(\tau^2-\rho\tilde{\rho})+(\tau^2-\rho\tilde{\rho}){\rm exp}[2i(\Delta+\Omega)L_{\rm arm}/c]}{1-\tilde{\rho} {\rm exp}[2i(\Delta+\Omega)L_{\rm arm}/c]}\\
&=\left[\frac{-\tilde{\rho}^*+{\rm exp}[2i(\Delta+\Omega)L_{\rm arm}/c]}{1-\tilde{\rho} {\rm exp}[2i(\Delta+\Omega)L_{\rm arm}/c]}\right](\tau^2-\rho\tilde{\rho}).
\end{split}
\ee
Note that $\tilde{\rho}=|\tilde{\rho}|e^{i{\rm Arg}[\rho]}$ and $|\tau^2-\rho\tilde{\rho}|=1$, therefore the above $r^{\rm idler}_{\rm ifo}$ can be written as:
\be
r^{\rm idler}_{\rm ifo}=\left[\frac{-|\tilde{\rho}|+{\rm exp}[2i(\Delta+\Omega)L_{\rm arm}/c+i{\rm Arg}[\rho]]}{1-|\tilde{\rho}| {\rm exp}[2i(\Delta+\Omega)L_{\rm arm}/c+i{\rm Arg}[\rho]]}\right]e^{i\phi_c},
\ee
where
\be
\phi_c={\rm Arg}\left[\tau^2-\rho\tilde{\rho}\right]-{\rm Arg}[\tilde{\rho}].
\ee
Since $\Omega L_{\rm SRC}/c\ll 1$, the $\phi_c$ dependence on $\Omega$ is very weak. Therefore this additional phase can be treated as almost a D-C phase. In our sample example, for compensating this additional phase angle, the phase of the homodyne detector of idler channel must be tuned by $\phi_c=-1.25$ rads.
 
\subsection{A sample rotation angle}
\begin{figure}[h]
   \centering
   \includegraphics[width=7in]{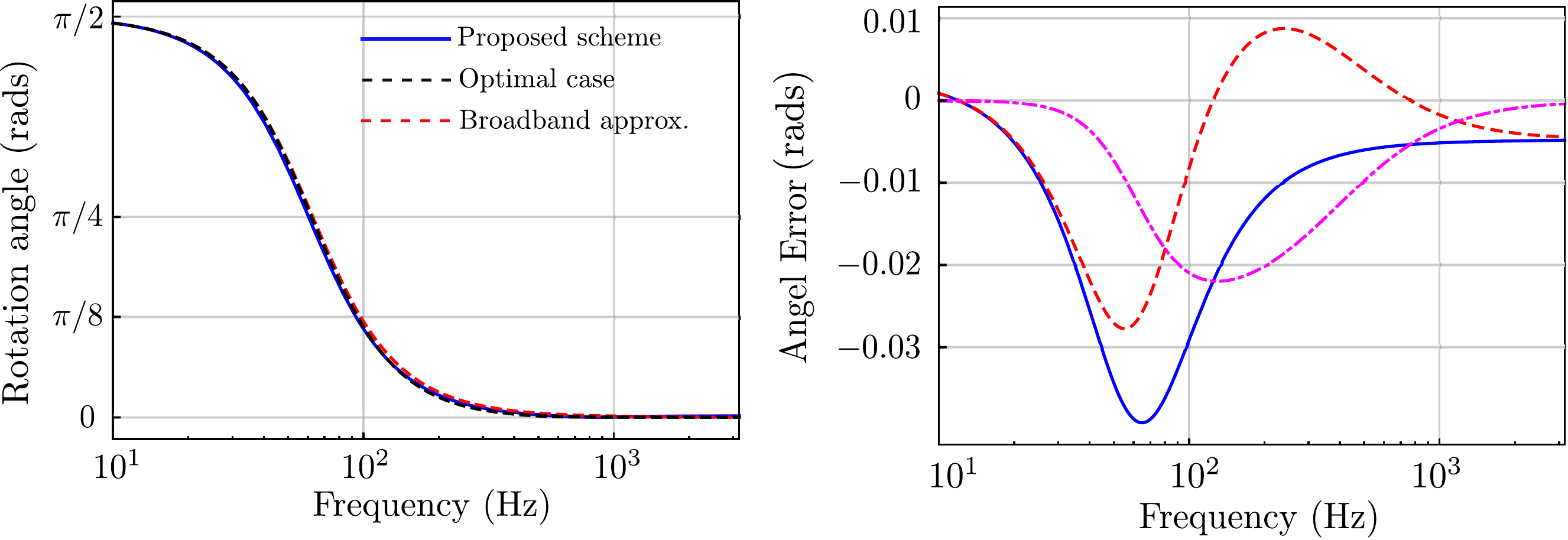} 
   \caption{Left panel: Rotation angle for the $\hat b-$ field. The blue curve is the result computed using the parameters given in Tab.1 of the main text and exact transfer matrix technique. The red dashed curve is the optimal result of $\Phi_{\rm rot}=-{\rm arctan}\mathcal{K}$ while the green dashed curve is the rotation angle when we approximate $\mathcal{K}$ using broadband approximation as discussed in the main text of this supplementary material. Right panel: Error of the rotation angle. (1)The magenta dotdashed curve is the difference between the analytical result of rotation angle under broadband approximation and the optimal angle ; (2) the blue curve is the difference between the rotation angle computed from our proposed scheme and the analytical result under broadband approximation; (3) the red dashed curve is the difference between the rotation angle computed from our proposed scheme and the optimal analytical result. }
   \label{fig:angle}
\end{figure}

Using the parameters given in Tab.\,\ref{tab:parameters}, we are able to produce the frequency dependent rotation angle for the $\hat b-$fields measured by the homodyne detector with local oscillator frequency $\omega_0+\Delta$,  as shown in Fig.\,\ref{fig:angle}.  This figure demonstrates that the proposed parameters in Tab.\,\ref{tab:parameters} lead to a result that is very close to the required rotation angle.
The angle error is also given in the right panel of Fig.\,\ref{fig:angle}, which shows that our result has maximally 0.04\,rads angle error in the intermediate frequency band (50-300\,Hz), creates a $40\%$ relative correction to the noise spectrum according to our estimation formula Eq.\,\eqref{eq:parameter_tolerance}, degrading the improvement factor from 12\,dB to 10.5\,dB. Exact computation shows that the degraded improvement factor is around 11.1\,dB.
\begin{table}[t]
   \centering
    \begin{threeparttable}
   \begin{tabular}{ |l|l|l| }
   \hline 
  $\lambda$ & carrier laser wavelength &1064nm\\
  $T_{\rm SRM}$ & SRM power transmissivity & 0.35\\
   $T$ & ITM power transmissivity & 0.014\\
  $L^{(0)}_{\rm arm}$ & Arm cavity initial length & 4km\tnote{1}\\
  $L^{(0)}_{\rm SRC}$& Signal recycling cavity initial length &50m\tnote{1}\\
  $\gamma$ & Detection bandwidth & 389 Hz\\
  $m$ & Mirror mass (ITM and ETM) & 40kg\\
  $I_c$ & Intra cavity power & 650kW\\
  $\Delta$ & Idler detuning & -300\,kHz-5${\rm FSR}_{SRC}$(=-15.3\,MHz)\\
  $r$ & Squeezing factor of the OPA & 1.23 (15dB)\\
  $\delta L_{\rm arm}$ & Arm length tuning & 19850$\lambda$\\
  $\delta L_{\rm src}$ & SRC length tuning & 26$\lambda$\\
  $\phi_c$& Phase compensation & -1.25 rads\\
  \hline
  \end{tabular}
   \begin{tablenotes}
        \footnotesize
        \item[1] These numbers are approximated value since the exact length should be integer number of half wavelength since both the arm cavity and signal recycling cavity should be on resonance with the main carrier light. In particular, the exact length of arm cavity closest to 4\,km is 3759398496$\lambda$; for exact length of signal recycling cavity closest to 50\,m is 46992481$\lambda$.
      \end{tablenotes}
  \caption{Sample Parameters for implementing our scheme on the Advanced LIGO. Here ${\rm FSR}_{\rm SRC}=c/(2L_{\rm SRC})$.}
   \label{tab:parameters}
   \end{threeparttable}
\end{table}

\section{Loss analysis}
Fig.3 of the main text takes into account of the loss in our system. There are four main loss sources in our design:  (1) the loss due to the arm cavity and signal recycling cavity, which currently has the value round $100$\,ppm (per round trip)  and $2000$\,ppm (per round trip) and has a small effect on the noise, compare to the current filter cavity design which has the value around $1$ppm per meter. (2) the input loss comes from the loss of the optical devices in the input optical path and mode mismatch and the readout loss comes from the measurement channel due to the non-perfect quantum efficiency of the photo-detector and the lossy optical devices in the output optical path and also mode mismatch. (3) The phase fluctuation of the local oscillators which are used to measure the $\hat a$ and $\hat b-$fields.

\subsection{Arm cavity loss and signal recycling cavity loss}
Similar to what has been discussed in\,\cite{Barsotti2014},  for the signal channel, $100$\,ppm round trip loss in the arm cavity corresponds to about 0.3\% total loss (considering the circulation of light fields) in advanced LIGO since it works in the resonant sideband extraction mode, which is comparabl to the impact of signal recycling cavity loss ($\sim 0.2\%$) at the interesting frequency band. However, since the large detuned idler field does not resonant with the signal recycling cavity, thereby a more careful simulation is needed. We simulate the effects of these noises in the following Fig.\,\ref{fig:losstransfer} to compare the impact of these different noise sources (a similar figure was also shown in\,\cite{Barsotti2014}). Note that at low frequency, the impact of these noises on the idler channel is much less than that on the signal channel, since idler field does not drive the interferometer mirrors through radiation pressure force noise in the interested frequency domain.
 
\begin{figure}[h]
   \centering
   \includegraphics[width=6in]{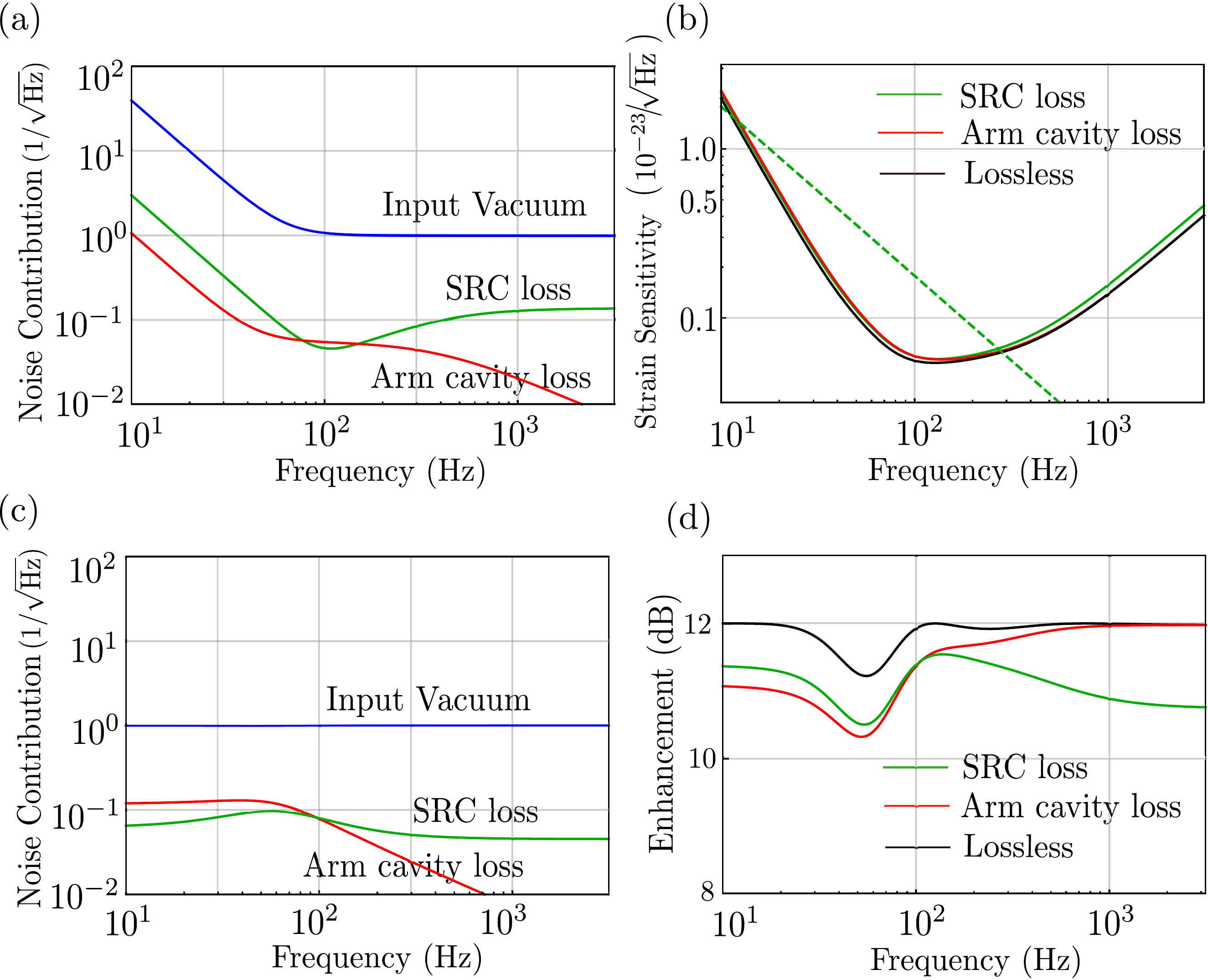} 
   \caption{Simulation of quantum noise contributions from the arm and
signal recycling cavities (round trip losses $\epsilon_c=100$ ppm and $\epsilon_{\rm SRC}=2000$ ppm respectively, as in the main text) and their impact on strain sensitivity and sensitivity enhancement. (a) Signal channel: comparison of quantum noise contributions in the signal channel from (unsqueezed) vacuum fluctuations, arm cavity loss, and SRC loss. The noise fields beat with a strong carrier field, resulting in radiation pressure noise below 50\,Hz due to the ponderomotive effect. (b) Idler channel: comparison of contributions in the idler channel from (unsqueezed) vacuum fluctuations, arm cavity loss, and SRC loss. The noise fields do not beat with a strong carrier field, and therefore do not participate in the ponderomotive process. (c) The (small) impact of loss due to combined signal and idler channels, shown separately for arm cavity loss and SRC loss (d) Impact of cavity losses on sensitivity
enhancement due to the proposed conditioning scheme.}
   \label{fig:losstransfer}
\end{figure}

\subsection{Input Loss and readout loss}
\begin{figure}[h]
   \centering
   \includegraphics[width=6in]{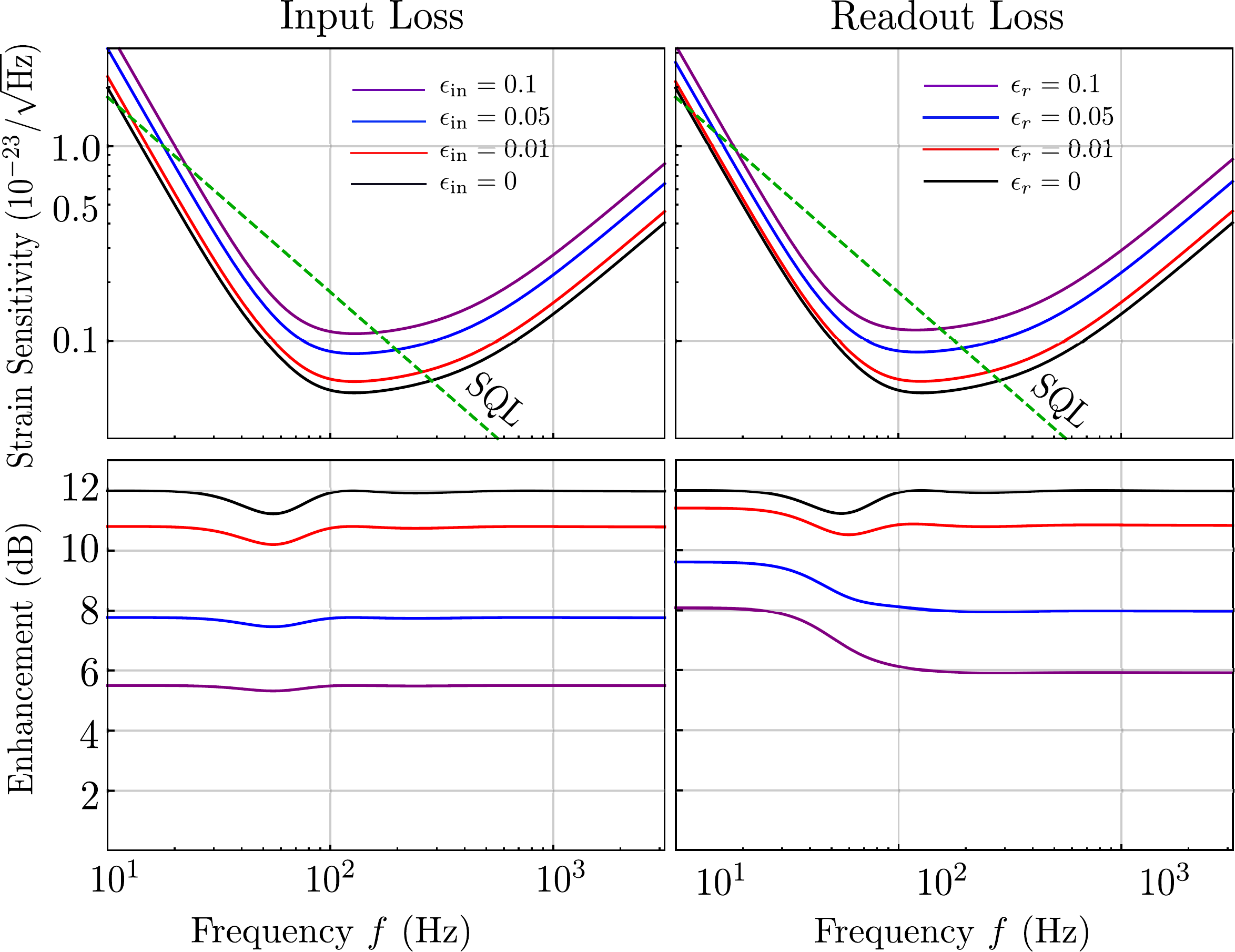} 
   \caption{Left panel: the strain sensitivity and enhancement factor of our configuration when there is only input loss. Right panel: the strain sensitivity and enhancement factor of our configuration when there is only readout loss. }
   \label{fig:loss_details}
\end{figure}

Let us now investigate the effect of the input loss and the readout loss.  The sensitivity curve in Fig.\,3 of the main text is computed using numerical transfer-matrix approach\,\cite{Corbitt2005}. Here for giving an analytical formula, we apply the single-mode approximation which is a very good approximation within one free-spectral-range of the arm cavity. The exact results about the contribution from the input loss and readout loss respectively are shown in Fig.\,\ref{fig:loss_details}.
Since the $\hat a$ and $\hat b$ fields propagate in a collinear way and share the same optical mode, therefore the input and readout loss of the $\hat a$ and $\hat b-$fields are the same, denoted by $\epsilon_{\rm in}$ (in terms of power) and we have:
\begin{align}
\hat a\rightarrow \sqrt{1-\epsilon_{\rm in}}\hat a+\sqrt{\epsilon_{\rm in}}\hat n^{\rm in}_{s},\quad
\hat b\rightarrow\sqrt{1-\epsilon_{\rm in}}\hat b+\sqrt{\epsilon_{\rm in}}\hat n^{\rm in}_i,\\
\hat A_2\rightarrow \sqrt{1-\epsilon_r}\hat A_2+\sqrt{\epsilon_r}\hat n^r_{s},\quad
\hat B_2\rightarrow\sqrt{1-\epsilon_r}\hat B_2+\sqrt{\epsilon_r}\hat n^r_i.
\end{align}
where $\hat n^{\rm in}_s$ and $\hat n^{\rm in}_i$ are two uncorrelated injection noises and $\hat n^r_s$ and $\hat n^r_i$ are two uncorrelated readout noises associated with two homodyne detectors. 

Expand to the first order of $\epsilon_{\rm in}$ and $\epsilon_{r}$, we have the approximated formula for the degradation of the strain sensitivity as a summation of input loss contribution $\Delta S^{\epsilon_{\rm in} \rm cond}_{\rm hh}$ and the readout loss contribution  $\Delta S^{\epsilon_r \rm cond}_{\rm hh}$:
\be
\Delta S_{\rm hh}^{\epsilon\rm cond}=\Delta S^{\epsilon_{\rm in} \rm cond}_{\rm hh}+\Delta S^{\epsilon_r \rm cond}_{\rm hh}
\ee
where:
\begin{align}
&\Delta S^{\epsilon_{\rm in} \rm cond}_{\rm hh}\approx\frac{h_{\rm SQL}^2}{2\cosh{2r}}\left(\mathcal{K}+\frac{1}{\mathcal{K}}\right)\left(\frac{2\cosh{2r}^2-\cosh{2r}-1}{\cosh{2r}}\right)\epsilon_{\rm in}\\
&\Delta S^{\epsilon_{r} \rm cond}_{\rm hh}\approx\frac{h_{\rm SQL}^2}{2}\left(\mathcal{K}\tanh^2{2r}+\frac{1+\tanh^2{2r}}{\mathcal{K}}\right)\epsilon_r.\\
\end{align}
It is easy to see that the effect of the input loss contributes a broadband degradation of squeezing degree. This degradation is frequency independent. However, for the readout loss, at high frequencies where $\mathcal{K}\ll1$ the relative loss correction is roughly $\Delta S^{\epsilon_{r} \rm cond}_{\rm hh}/S_{\rm hh}^{\rm cond}\approx [1+(\tanh{2r})^2]\epsilon_r$, while at low frequencies where $\mathcal{K}\gg 1$, $\Delta S^{\epsilon_{r} \rm cond}_{\rm hh}/S_{\rm hh}^{\rm cond}$ is roughly $(\tanh{2r})^2\epsilon_r$. Therefore the readout loss effect at high frequencies is higher than that at low frequencies,  due to the fact that the pondermotive effect amplifies the signal at the low frequency band\,\cite{Kimble2001}. This character of the sensitivity curves has been shown in Fig.\,3 of the main text and more explicitly in Fig.\,\ref{fig:loss_details} of this Supplementary Material.

If we take the assumption of $\epsilon_{\rm in}=\epsilon_{r}=\epsilon$, the noise spectrum for the traditional broadband squeezing using an additional filter cavity and the conditional squeezing, including the loss $\epsilon$ to the first order, and under the approximation that the squeezing degree is large, we have:
\be
\Delta S_{\rm hh}^{\epsilon\rm cond}\approx\frac{h_{\rm SQL}^2}{2}\left(\frac{2}{\mathcal{K}}+\frac{3}{2}\mathcal{K}\right)2\epsilon,\quad \Delta S_{\rm hh}^{\epsilon\rm tran}\approx\frac{h_{\rm SQL}^2}{2}\left(\frac{2}{\mathcal{K}}+\mathcal{K}\right)\epsilon,
\ee
where $\Delta S_{\rm hh}^{\epsilon\rm tran}$ is the first order correction to the sensitivity curve produced by traditional squeezing scheme\,\cite{Kimble2001,Miao2014} and clearly we have $\Delta S_{\rm hh}^{\epsilon\rm cond}\approx2\Delta S_{\rm hh}^{\epsilon\rm tran}$. As we have briefly mentioned in the main text, due to the fact that both signal and idler beams experience the same loss during their propagation in our scheme, the input and readout losses in our configuration are roughly doubled compare to that of the traditional squeezing scheme with an additional filter cavity.

\subsection{Phase fluctuation}
The classical phase uncertainty of local oscillators in homodyne detection scheme has a very small effect on the sensitivity, proved as follows.

We assume that the phase fluctuation of the local oscillators used for signal and idler detection are independent Gaussian random variables represented by $\xi_s$ and $\xi_i$ respectively, with zero mean and standard variacne $\xi_{vs}$ and $\xi_{vi}$, such that their probability density functions are given by:
\be
P_s(\xi_s)=\frac{1}{\sqrt{2\pi\xi_{vs}^2}}e^{-\frac{\xi_s^2}{2\xi_{vs}^2}},\quad
P_i(\xi_i)=\frac{1}{\sqrt{2\pi\xi_{vi}^2}}e^{-\frac{\xi_i^2}{2\xi_{vi}^2}},
\ee
For measuring any field $\hat O$, the effect of phase fluctuation $\xi_O$ in a phase quadrature readout scheme is that the actual detected quadrature is also a random variable which is given by
$\hat O_m(t)=-\hat O_1(t)\sin{\xi_O}+\hat O_2(t)\cos{\xi_O}$. The ensemble-averaged variance of the detected quadrature is then\,\cite{Aoki2006}:
\be
\begin{split}
S_{\hat O_m\hat O_m}&=\frac{1}{2}\int d\xi_O P(\xi_O)\langle \hat O_m \hat O_m^\dag+ \hat O_m^\dag\hat O_m\rangle\\&=V_{\hat O_1\hat O_1} \int d\xi_OP(\xi_O)\sin\xi_O^2+V_{\hat O_2\hat O_2} \int d\xi_OP(\xi_O)\cos\xi_O^2+(V_{\hat O_1\hat O_2}+V_{\hat O_2\hat O_1})\sin\xi_O\cos\xi_O\\
\end{split}
\ee
Now if we use the identities (we assume that $\xi_{vo}\ll1$.):
\be
\begin{split}
&\int d \xi_O P(\xi_O)\sin^2\xi_O=e^{-\xi_{ov}^2}\sinh{\xi_{ov}^2}\approx \xi_{ov}^2\approx \sin{\xi_{ov}^2}\\
&\int d \xi_O P(\xi_O)\cos^2\xi_O=e^{-\xi_{ov}^2}\cosh{\xi_{ov}^2}\approx 1-\xi_{ov}^2\approx\cos{\xi_{ov}^2},
\end{split}
\ee
and the fact that the $\sin\xi_O\cos\xi_O$ is an odd function, one can obtain:
\be\label{phase_jittering_variance}
S_{\hat O_m\hat O_m}\approx \cos{\xi_{vo}^2}V_{\hat O_1\hat O_1}+\sin{\xi_{vo}^2}V_{\hat O_2\hat O_2}.
\ee
Using the above formula, one can can compute the variance of signal and idler fields.

Similarly, the cross correlation between the signal and idler fields (represented by $\hat A$ and $\hat B$) in the phase quadrature readout scheme, accounting for phase fluctuations, is given by:
\be
S_{A_mB_m}=\frac{1}{2}\int d\xi_id\xi_s P(\xi_i)P(\xi_s)\langle(\hat A_1\sin\xi_s+\hat A_2\cos\xi_s)(\hat B_1\cos\xi_i+\hat B_2\sin\xi_i)\rangle.
\ee
Because $\xi_i$ and $\xi_s$ are two independent random variables, the above cross correlation fluctuation leads to:
\be\label{phase_jittering_cross}
S_{A_mB_m}=S_{A_2B_1}e^{-\frac{\xi_{vs}^2+\xi_{vi}^2}{2}}\approx S_{a_2b_1}\left(1-\frac{\xi_{vs}^2+\xi_{vi}^2}{2}\right).
\ee
where we have used the identities:
\be
\int d \xi_O P(\xi_O)\cos\xi_O=e^{-\frac{\xi_O^2}{2}},\quad\int d \xi_O P(\xi_O)\sin\xi_O=0.
\ee
Substituting Eq.\,\eqref{phase_jittering_variance} and Eq.\,\eqref{phase_jittering_cross} into Eq.\,\eqref{eq:opt_var}, the final strain sensitivity can be written as:
\be
\begin{split}
S_{hh}=\frac{h_{\rm SQL}^2}{2\cosh{2r}}\left[\frac{1+(\xi_{vs}^2+\xi_{vi}^2)\sinh{4r}}{\mathcal{K}}+(1-\xi_{vs}^2+\xi_{vi}^2\sinh{4r})\mathcal{K}\right].
\end{split}
\ee
The typical experimental rms of the local oscillator phase is taken to be $1.7$m rad such as shown in\,\cite{Vahlbruch2016}, that means the phase fluctuation quantities in the above formula have an orders of magnitude $\sim 10^{-6}$ rad$^2$. For a 15dB squeezer, $1$\,mrad phase jittering only contributes roughly $\sim 0.5\%$ relative correction to the final sensitivity.

\bibliographystyle{unsrt}

\begin{thebibliography}{10}

\bibitem{GW150914}
B.~Abbott~(et al.).
\newblock Observation of {G}ravitational {W}aves from a {B}inary {B}lack {H}ole
  {M}erger.
\newblock {\em Phys. Rev. Lett}, 116(061102), 2016.

\bibitem{Punturo2010}
M.~Punturo~(et al.).
\newblock The {E}instein {T}elescope: a third-generation gravitational wave
  observatory.
\newblock {\em Classical and Quantum Gravity}, 27(19), 2010.

\bibitem{Danilishin2012}
S.~Danilishin and F.~Ya. Khalili.
\newblock Quantum {M}easurement {T}heory in {G}ravitational-{W}ave {D}etectors.
\newblock {\em Living Rev. Relativity}, 15(5), 2012.

\bibitem{Miao2014}
H.~Miao, H.~Yang, R.~X. Adhikari, and Y.~Chen.
\newblock Quantum limits of interferometer topologies for gravitational
  radiation detection.
\newblock {\em Classical Quantum Gravity}, 31(16), August 2014.

\bibitem{Dwyer2015}
S.~Dwyer, D.~Sigg, S.~Ballmer, L.~Barsotti, N.~Mavalvala, and M.~Evans.
\newblock Gravitational wave detector with cosmological reach.
\newblock {\em Phys. Rev. D}, 91(082001), 2015.

\bibitem{GWsensitivity}
B.~Abbott~(et al.).
\newblock Exploring the sensitivity of next generation gravitational wave
  detectors.
\newblock {\em LIGO Document}, (P1600143-v14), 2016.

\bibitem{Sathyaprakash2012}
B.~Sathyaprakash~(et. al).
\newblock Scientific objectives of {E}instein {T}elescope.
\newblock {\em Classical and Quantum Gravity}, 29(12), 2012.

\bibitem{Tso2016}
R.~Tso, M.~Isi, Y.~Chen, and L.~Stein.
\newblock Modeling the dispersion and polarization content of gravitational
  waves for tests of general relativity.
\newblock In {\em Seventh Meeting on CPT and Lorentz Symmetry}, June 2016.

\bibitem{Kostelecky2016}
A.~Kostelecky and M.~Mewes.
\newblock Testing local {L}orentz invariance with gravitational waves.
\newblock {\em Phys. Lett. B}, 757:510--514, 2016.

\bibitem{Lasky2016}
Lasky, E.~Thrane, Yu. Levin, J.~Blackman, and Y.~Chen.
\newblock Detecting gravitational-wave memory with ligo: implications of
  gw150914.
\newblock {\em Phys. Rev. Lett}, 117(061102), 2016.

\bibitem{Berti2013}
E.~Berti.
\newblock Astrophysical {B}lack {H}oles as {N}atural {L}aboratories for
  {F}undamental {P}hysics and {S}trong-{F}ield {G}ravity.
\newblock {\em Braz J Phys}, 43(341), 2013.

\bibitem{Chu2016}
Q.~Chu, E.~J. Howell, A.~Rowlinson, H.~Gao, B.~Zhang, S.~J. Tingay,
  M.~Bo{\"e}r, and L.~Wen.
\newblock Capturing the electromagnetic counterparts of binary neutron star
  mergers through low-latency gravitational wave triggers.
\newblock {\em MNRAS}, 459:121--139, Jun 2016.

\bibitem{Metzger2012}
B.~D. Metzger and E.~Berger.
\newblock What is the most promising electromagnetic counterpart of a neutron
  star binary merger?
\newblock {\em The Astrophysical Journal}, 746(1):48, 2012.

\bibitem{cosmicstring2014}
J.~Aasi(et.~al).
\newblock Constraints on {C}osmic {S}trings from the {LIGO}-{V}irgo
  {G}ravitational-{W}ave {D}etectors.
\newblock {\em Phys. Rev. Lett}, 112(131101), 2014.

\bibitem{Drever1976}
R.~W.~P. Drever, J.~Hough, W.~A. Edelstein, J.~R. Pugh, and W.~Martin.
\newblock In B.~Bertotii, editor, {\em Experimental Gravitation}, page 365,
  Pavia, Italy., 1976. Accademia Nazionale dei Lincei.

\bibitem{Braginsky1977}
V.~B Braginsky and Yu.~I. Vorontsov.
\newblock {\em Usp. Fiz. Nauk}, 114(41), 1974.

\bibitem{Weiss1979}
R.~Weiss.
\newblock {\em Sources of Gravitational Radiation}.
\newblock Cambridge University Press, 1979.

\bibitem{Caves1981}
C.~M. Caves.
\newblock Quantum-mechanical noise in an interferometer.
\newblock {\em Phys. Rev. D}, 23(1693), April 1981.

\bibitem{Kimble2001}
H.~J. Kimble, Yu. Levin, A.~B. Matsko, K.~S. Thorne, and S.~P. Vyatchanin.
\newblock Conversion of conventional gravitational-wave interferometers into
  quantum nondemolition interferometers by modifying their input and/or output
  optics.
\newblock {\em Phys. Rev. D}, 65(022002), 2001.

\bibitem{Buonanno2001}
A.~Buonanno and Y.~Chen.
\newblock Quantum noise in second generation, signal-recycled laser
  interferometric gravitational-wave detectors.
\newblock {\em Phys. Rev. D}, 64(042006), 2001.

\bibitem{Buonanno2003}
A.~Buonanno and Y.~Chen.
\newblock Scaling law in signal recycled laser-interferometer
  gravitational-wave detectors.
\newblock {\em Phys. Rev. D}, 67(062002), 2003.

\bibitem{Unruh1982}
W.~G Unruh.
\newblock {\em Quantum Noise in the Interferometer Detector}, page 647.
\newblock Plenum Press, New York, 1983.

\bibitem{Schnabel2010}
R.~Schnabel, N.~Mavalvala, D.~E. McClelland, and P.~K. Lam.
\newblock Quantum metrology for gravitational wave astronomy.
\newblock {\em Nat. Commun.}, 1(121), 2010.

\bibitem{LIGOwhitepaper}
{LIGO} {I}nstrument {S}cience {W}hite {P}aper.
\newblock Technical report, Feb. 2015.

\bibitem{Vahlbruch2008}
H.~Vahlbruch, M.~Mehmet, S.~Chelkowski, B.~Hage, A.~Franzen, N.~Lastzka,
  S.~Gossler, K.~Danzmann, and R.~Schnabel.
\newblock Observation of {S}queezed {L}ight with 10-d{B} {Q}uantum-{N}oise
  {R}eduction.
\newblock {\em Phys. Rev. Lett}, 100(033602), 2009.

\bibitem{Mehmet2011}
M~Mehmet, S.~Ast, T.~Eberle, S.~Steinlechner, H.~Vahlbruch, and R.~Schnabel.
\newblock Squeezed light at 1550 nm with a quantum noise reduction of 12.3
  d{B}.
\newblock {\em Optics Express}, 19(25763), 2011.

\bibitem{Chua2011}
S.~S.~Y. Chua, M.~S. Stefszky, C.~M. Mow-Lowry, B.~C. Buchler, S.~Dwyer, D.~A.
  Shaddock, P.~K. Lam, and D.~E. McClelland.
\newblock Backscatter tolerant squeezed light source for advanced
  gravitational-wave detectors.
\newblock {\em Optics Letters}, 36(23):4680--4682, 2011.

\bibitem{Stefszky2012}
M.~S. Stefszky, C.~M. Mow-Lowry, S.~S.~Y. Chua, D.~A. Shaddock, B.~C. Buchler,
  H.~Vahlbruch, A.~Khalaidovski, R.~Schnabel, P.~K. Lam, and D.~E. McClelland.
\newblock Balanced homodyne detection of optical quantum states at audio-band
  frequencies and below.
\newblock {\em Classical Quantum Gravity}, 29(145015), 2012.

\bibitem{McKenzie2004}
K.~McKenzie, N.~Grosse, W.~P. Bowen, S.~E. Whitcomb, M.~B. Gray, D.~E.
  McClelland, and P.~K. Lam.
\newblock Squeezing in the {A}udio {G}ravitational-{W}ave {D}etection {B}and.
\newblock {\em Phys. Rev. Lett}, 93(161105), 2004.

\bibitem{Vahlbruch2016}
H.~Vahlbruch, M.~Mehmet, K.~Danzmann, and R.~Schna\-bel.
\newblock Detection of 15 d{B} squeezed states of light and their application
  for the absolute calibration of photo-electric quantum efficiency.
\newblock {\em Phys. Rev. Lett (accepted)}, 2016.

\bibitem{squeezingLIGO2011}
B.~Abbott~(et al.).
\newblock A gravitational wave observatory operating beyond the quantum
  shot-noise limit.
\newblock {\em Nature Physics}, 7(962), 2011.

\bibitem{squeezingLIGO2013}
B.~Abbott~(et al.).
\newblock Enhanced sensitivity of the {LIGO} gravitational wave detector by
  using squeezed states of light.
\newblock {\em Nature Photonics}, 7(613), 2013.

\bibitem{Jaekel1990}
M.~T. Jaekel and S.~Reynaud.
\newblock Quantum {L}imits in {I}nterferometer {M}easurement.
\newblock {\em EPL (Europhysics Letters)}, 13(4):301, 1990.

\bibitem{Chelkowski2005}
S.~Chelkowski, H.~Vahlbruch, B.~Hage, A.~Franzen, N.~Lastzka, K.~Danzmann, and
  R.~Schnabel.
\newblock Experimental characterization of frequency-dependent squeezed light.
\newblock {\em Phys. Rev. A}, 71(013806), 2005.

\bibitem{Evans2013}
M.~Evans, L.~Barsotti, P.~Kwee, J.~Harms, and H.~Miao.
\newblock Realistic filter cavities for advanced gravitational wave detectors.
\newblock {\em Phys. Rev. D}, 88(022002), 2013.

\bibitem{Khalili2007}
F.~Ya. Khalili.
\newblock Quantum variational measurement in the next generation
  gravitational-wave detectors.
\newblock {\em Phys. Rev. D}, 76(102002), 2007.

\bibitem{Khalili2010}
F.~Ya. Khalili.
\newblock Optimal configurations of filter cavity in future gravitational-wave
  detectors.
\newblock {\em Phys. Rev. D}, 81(122002), 2010.

\bibitem{Isogai2013}
T.~Isogai, J.~Miller, P.~Kwee, L.~Barsotti, and M.~Evans.
\newblock Loss in long-storage-time optical cavities.
\newblock {\em Optics Express}, 21(24):30114--30125, 2013.

\bibitem{Kwee2014}
P.~Kwee, J.~Miller, T.~Isogai, L.~Barsotti, and M.~Evans.
\newblock Decoherence and degradation of squeezed states in quantum filter
  cavities.
\newblock {\em Phys. Rev. D}, 90(062006), 2014.

\bibitem{Caposcasa2016}
E~Caposcasa, M~Barsuglia, J~Degallaix, L~Pinard, N.~Straniero, R.~Schnabel,
  K.~Somiya, Y.~Aso, D.~Tatsumi, and R.~Flaminio.
\newblock Estimation of losses in a 300 m filter cavity and quantum noise
  reduction in the kagra gravitational-wave detector.
\newblock {\em Phys. Rev. D}, 93(082004), 2016.

\bibitem{ET2011}
{ET}~{Science} {Team}.
\newblock Einstein gravitative wave telescope conceptual design study.
\newblock {\em ET-0106C-10}, 2011.

\bibitem{Mikhailov2006}
E.~E. Mikhailov, K.~Goda, T.~Corbitt, and N.~Mavalvala.
\newblock Frequency-dependent squeeze-amplitude attenuation and squeeze-angle
  rotation by electromagnetically induced transparency for gravitational-wave
  interferometers.
\newblock {\em Phys. Rev. A}, 73(053810), 2006.

\bibitem{Ma2014}
Y.~Ma, S.~L. Danilishin, C.~Zhao, H.~Miao, W.~Z. Korth, Y.~Chen, R.~L. Ward,
  and D.~G. Blair.
\newblock Narrowing the {F}ilter-{C}avity {B}andwidth in {G}ravitational-{W}ave
  {D}etectors via {O}ptomechanical {I}nteraction.
\newblock {\em Phys. Rev. Lett}, 113(151102), 2014.

\bibitem{Qin2014}
J~Qin, C~Zhao, Y~Ma, X~Chen, L~Ju, and D.~G. Blair.
\newblock Classical demonstration of frequency-dependent noise ellipse rotation
  using optomechanically induced transparency.
\newblock {\em Phys. Rev. A}, 89(041802(R)), 2014.

\bibitem{Zhang2003}
J.~Zhang.
\newblock {E}instein-{P}odolsky-{R}osen sideband entanglement in broadband
  squeezed light.
\newblock {\em Phys. Rev. A}, 67(054302), 2003.

\bibitem{Marino2007}
A.~M Marino, C.~R Stroud, V~Wong, R.~S Bennink, and R.~W. Boyd.
\newblock Bichromatic local oscillator for detection of two-mode squeezed
  states of light.
\newblock {\em J. Opt. Soc. Am. B}, 24(2):335--339, 2007.

\bibitem{Hage2010}
B.~Hage, A.~Samblowski, and R.~Schnabel.
\newblock Towards {E}instein-{P}odolsky-{R}osen quantum channel multiplexing.
\newblock {\em Phys. Rev. A}, 81(062301), June 2010.

\bibitem{Barsotti2016}
L.~Barsotti.
\newblock Ligo: {T}he {A}+ {U}pgrade.
\newblock {\em LIGO Document}, (G1601199-v2), 2016.

\bibitem{Barsotti2014}
L.~Barsotti.
\newblock Squeezing for {A}dvanced {LIGO}.
\newblock {\em LIGO Document}, G1401092-v1, 2014.


\end{thebibliography}

\begin{thebibliography}{10}

\bibitem{Scully1997}
M.~O. Scully and M.~S. Zubairy.
\newblock {\em {Q}uantum {O}ptics ({C}hapter 16)}.
\newblock Cambridge University Press, 1997.

\bibitem{Kimble2001}
H.~J. Kimble, Yu. Levin, A.~B. Matsko, K.~S. Thorne, and S.~P. Vyatchanin.
\newblock Conversion of conventional gravitational-wave interferometers into
  quantum nondemolition interferometers by modifying their input and/or output
  optics.
\newblock {\em Phys. Rev. D}, 65(022002), 2001.

\bibitem{Purdue2002}
P.~Purdue and Y.~Chen.
\newblock Practical speed meter designs for quantum nondemolition
  gravitational-wave interferometers.
\newblock {\em Phys. Rev. D}, 66(122004), 2002.

\bibitem{Khalili2007}
F.~Ya. Khalili.
\newblock Quantum variational measurement in the next generation
  gravitational-wave detectors.
\newblock {\em Phys. Rev. D}, 76(102002), 2007.

\bibitem{Buonanno2003}
A.~Buonanno and Y.~Chen.
\newblock Scaling law in signal recycled laser-interferometer
  gravitational-wave detectors.
\newblock {\em Phys. Rev. D}, 67(062002), 2003.

\bibitem{Barsotti2014}
L.~Barsotti.
\newblock Squeezing for {A}dvanced {LIGO}.
\newblock {\em LIGO Document}, G1401092-v1, 2014.

\bibitem{Miao2014}
H.~Miao, H.~Yang, R.~X. Adhikari, and Y.~Chen.
\newblock Quantum limits of interferometer topologies for gravitational
  radiation detection.
\newblock {\em Classical Quantum Gravity}, 31(16), August 2014.

\bibitem{Corbitt2005}
T.~Corbitt, Y.~Chen, and N.~Mavalvala.
\newblock Mathematical framework for simulation of quantum fields in complex
  interferometers using the two-photon formalism.
\newblock {\em Phys. Rev. A}, 72(013818), July 2005.

\bibitem{Aoki2006}
T.~Aoki, G.~Takahashi, and A~Furusawa.
\newblock Squeezing at 946nm with periodically poled {KTiOPO4}.
\newblock {\em Opt. Express}, 14(15):6930--6935, 2006.

\bibitem{Vahlbruch2016}
H.~Vahlbruch, M.~Mehmet, K.~Danzmann, and R.~Schnabel.
\newblock Detection of 15 db squeezed states of light and their application for
  the absolute calibration of photo-electric quantum efficiency.
\newblock {\em Phys. Rev. Lett (accepted)}, 2016.

\end{thebibliography}

\end{document}